\documentclass[aps,twocolumn,floatfix]{revtex4} 

\usepackage{amsmath,amsfonts,amssymb,graphicx,color,times,bbm}

\begin{document}

\bibliographystyle{apsrev}
\newtheorem{theorem}{Theorem}
\newtheorem{proposition}{Proposition}
\newtheorem{lemma}{Lemma}
\newcommand{\proofend}{\hfill\fbox\\\medskip }
\newcommand{\proof}[1]{{\bf Proof.} #1 $\proofend$}
\newcommand{\nn}{{\mathbbm{N}}}
\newcommand{\rr}{{\mathbbm{R}}}
\newcommand{\cc}{{\mathbbm{C}}}
\newcommand{\mbp}{\ensuremath{\spadesuit}}
\newcommand{\je}{\ensuremath{\heartsuit}}
\newcommand{\jd}{\ensuremath{\clubsuit}}
\newcommand{\id}{{\mathbbm{1}}}
\renewcommand{\vec}[1]{\boldsymbol{#1}}
\newcommand{\me}{\mathrm{e}}
\newcommand{\mi}{\mathrm{i}}
\newcommand{\md}{\mathrm{d}}
\newcommand{\sg}{\text{sgn}}

\delimitershortfall=-2pt

\title{Statistics dependence of the entanglement entropy}

\author{M.\ Cramer$^{1,2,3}$, 
J.\ Eisert$^{2,3}$,  and M.B.\ Plenio$^{2,3}$}

\affiliation{1 Institut f{\"u}r Physik, Universit{\"a}t Potsdam,
Am Neuen Palais 10, D-14469 Potsdam, Germany\\
2 QOLS, Blackett Laboratory, Imperial College London,
Prince Consort Road, London SW7 2BW, UK\\
3 Institute for Mathematical Sciences,
Imperial College London, Exhibition Road, London, SW7 2BW, UK}

\begin{abstract}
The entanglement entropy of a distinguished region of a 
quantum many-body system reflects the entanglement present 
in its pure ground state. In this work, we establish scaling 
laws for this entanglement for {\it critical} quasi-free 
fermionic and bosonic lattice systems, without resorting 
to numerical means. We consider the geometrical setting 
of $D$-dimensional half-spaces which allows us to 
exploit a connection to the one-dimensional case.
Intriguingly, 
we find a difference in the scaling properties depending on 
whether the system is bosonic---where an area-law is first 
proven to hold---or fermionic, extending previous findings 
for cubic regions. For bosonic systems with nearest neighbor 
interaction we prove the conjectured area-law by computing 
the logarithmic negativity analytically. We identify a length scale associated 
with entanglement, different from the correlation length.
For fermions we determine the logarithmic 
correction to the area-law, which depends on the topology 
of the Fermi surface. We find that Lifshitz quantum phase 
transitions are accompanied with a non-analyticity in the 
prefactor of the leading order term.
\end{abstract}

\maketitle

\date{\today}


The occurrence of critical points at zero temperature 
holds the key to the understanding of several phenomena 
in quantum many-body systems in the condensed matter 
context \cite{Vojta}. Quantum criticality is accompanied
by a divergence of the typical length scale, the correlation
length. These long-range correlations come 
along with genuine entanglement in the ground state, grasped by the entanglement 
entropy $E_S = S( \text{tr}_{\backslash \mathcal{A}}[\rho])$. 
This is the entropy of the reduced density matrix that 
is obtained when tracing out the degrees of freedom outside
a distinguished region ${\cal A}$, quantifying
the degree of entanglement between this region and the 
rest [2--19].

This notion of the entanglement or geometric entropy---and more to the point its scaling behavior
abstracting from details of the model---has enjoyed a 
strong revival of interest recently, partially 
driven by intuition from quantum information theory: 
previously conjectured scaling laws in higher dimensions \cite{OldArea}, relating 
the entanglement entropy to the boundary area---not the volume---of the region, have been 
rigorously established using quantum information 
ideas [4--6]. This was followed 
by observations of violations of such area-laws 
\cite{Wolf}. The entanglement entropy has in its 
non-leading-order behavior interestingly been linked 
to the topology of the system \cite{Preskill}, using 
ideas of topological quantum field theory, and been 
studied under time evolution \cite{Fazio}. Partly, 
this renewed interest is triggered by the implications 
on the simulatability of quantum systems using 
density-matrix renormalization approaches: the 
entanglement entropy quantifies in a sense 
the relevant number of degrees of freedom to be 
considered \cite{Peschel1,DMRG}. 
 
If entanglement is to reflect critical 
or non-critical properties of quantum many-body 
systems, an area-relationship might of course be 
expected to hold or not, depending on whether the
two-point correlation functions diverge. One might 
be tempted to think that entanglement could yet be 
seen as an indicator of criticality in the same 
sense. Intriguingly, it  turns out that the situation 
is more complex. As we will also see,
even for critical systems, an area-relationship can
hold, despite a divergent correlation length (as 
can also be observed in projected entangled pair 
states, satisfying an area-law by construction \cite{PEPS}). 
In this work, we demonstrate with a fully analytical argument
that it can depend on the statistics of the system---bosonic or fermionic---whether an area-relationship holds
or is in fact violated. In this way, we resolve the key open
question: ``What happens in the critical bosonic case?''
In the process we confirm some conjectures based on 
numerical findings for small system size 
\cite{MustCite,OldArea,Roscilde} 
and refute others, such as the 
conjecture of a break-down of an area law for 
critical bosons in $D>1$ in Ref.\ \cite{Michael}.

\begin{figure}
\begin{center}
\includegraphics[width=0.9\columnwidth]{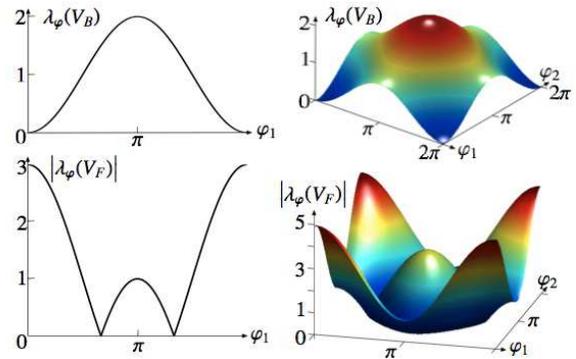}
\end{center}
\caption{\label{fig}Spectra of critical nearest-neighbour Hamiltonians (top: bosonic, bottom: fermionic) in one and two dimensions. The spectrum of the individual decoupled chains is given by the spectrum of the full Hamiltonian along the first coordinate $\varphi_1$.  
The topology of the set of solutions to 
$\lambda_{\vec{\varphi}}=0$---the Fermi-surface in the 
fermionic case---provides an intuition as to why the scaling behavior of entanglement is different for fermions then for bosons in $D>1$, see text.}
\end{figure}

Here, we establish first analytical scaling laws 
for critical bosonic systems. We achieve these results 
for the geometrical setting of a half-space in $D$-dimensions, 
completing the program initiated in Ref.\ \cite{Area}. 
These findings are 
compared with fermionic half-spaces, complementing recent
results on cubic regions in Refs.\ \cite{Wolf}, and 
in a fashion consistent with numerical work in 
Ref.\ \cite{MustCite,Roscilde}.
We treat bosonic and fermionic systems on the same footing --
in terms of Majorana operators for fermions and canonical
coordinates for bosons. We hence provide a
unified and complete framework for entanglement
scaling in critical quasi-free systems with that geometry.

{\em The setting. --} We consider cubic lattices of spatial dimension $D$, 
$\mathcal{L}=[1,\dots,N]^{\times D}$,  $|\mathcal{L}|=N^D$, equipped with 
{\it periodic boundary 
conditions} and study ground states $\varrho$ of 
Hamiltonians that are quadratic forms of either bosonic 
or fermionic operators. The geometric setting is that of 
a half-space, distinguishing w.l.o.g. the first spatial direction, considering 
a subsystem $\mathcal{A}= [1,\dots, M]\times \mathcal{L}'$,  
$\mathcal{L}'=[1,\dots,N]^{\times D-1}$, and its 
entanglement with the rest
$\mathcal{B}=\mathcal{L}
\backslash\mathcal{A}=[M+1,\dots,N]\times \mathcal{L}'$. We can hence make 
use of an idea of exploiting transverse momenta. This 
geometrical setting has notably numerically been assessed with 
respect to local spectra and simulatability issues in 
the seminal work Ref.\ \cite{Peschel1}.

When we say that (i) this entanglement as 
quantified by the
von-Neumann entropy $E_S(M,N)=S(\varrho_{\mathcal{A}})$ of the 
reduced state $\varrho_{\mathcal{A}}=\text{tr}_{\mathcal{B}}[\varrho]$ 
satisfies an {\it area-law}, we mean that for $M=N/2$ the 
entanglement entropy satisfies
\begin{equation}\nonumber
        \mathcal{E}:=\lim_{N\rightarrow\infty} E_S(M,N)/N^{D-1}\leq 
        \text{const},
\end{equation}  
i.e., it scales at most like the {\it boundary area} of 
$\mathcal{A}$.

For systems {\it violating} 
the area-law, $\mathcal{E} = \infty$, we will (ii) study the exact form of the encountered logarithmic divergence in $M$,
\begin{equation}\nonumber
        \mathcal{E} = \text{const}\times  \log M + o(\log M).
\end{equation}

We will subsequently discuss physical systems that are described 
by Hamiltonians of the type
\begin{eqnarray}\nonumber
        \hat{H}=\frac{1}{2}\sum_{\vec{i},\vec{j}}
        \left[
        \hat{d}_{\vec{i}}^\dagger A_{\vec{i},\vec{j}}\hat{d}_{\vec{j}}
        +\hat{d}_{\vec{i}} B_{\vec{i},\vec{j}}\hat{d}_{\vec{j}}^\dagger
        +\hat{d}_{\vec{i}} C_{\vec{i},\vec{j}}\hat{d}_{\vec{j}}
        +\hat{d}_{\vec{i}}^\dagger D_{\vec{i},\vec{j}}
        \hat{d}_{\vec{j}}^\dagger
        \right],
\end{eqnarray}
where operators $\hat{d}_{\vec{i}}$ are either bosonic or fermionic
and vectors $\vec{i}=(i_1\cdots i_D)\in \mathcal{L}$ label 
individual sites of the cubic lattice. To ensure hermiticity we 
demand the real coefficients to satisfy 
$A_{\vec{i},\vec{j}}=B_{\vec{i},\vec{j}}=A_{\vec{j},\vec{i}}$ and 
$C_{\vec{i},\vec{j}}=D_{\vec{i},\vec{j}}=C_{\vec{j},\vec{i}}$ for bosons, and 
$A_{\vec{i},\vec{j}}=-B_{\vec{i},\vec{j}}=A_{\vec{j},\vec{i}}$, 
$C_{\vec{i},\vec{j}}=-D_{\vec{i},\vec{j}}=-C_{\vec{j},\vec{i}}$  for 
fermions. Furthermore, we assume translational invariance and 
periodic boundary conditions (all coupling matrices depend only 
on the difference $\vec{i}-\vec{j}$ and are cyclic matrices).  We will 
lead the discussion in terms of hermitian operators 
$\hat{\vec{r}}=(\hat{x}_1,\dots,\hat{x}_{|\mathcal{L}|}, \hat{p}_1,\dots
,\hat{p}_{|\mathcal{L}|})^t$ (in a mild abuse of notation, the transposition refers to the tuple, not to operators) defined by 
$\hat{x}_{\vec{i}}=({\hat{d}_{\vec{i}}+\hat{d}_{\vec{i}}^\dagger})/\sqrt{2} $ and
        $\hat{p}_{\vec{i}}=-\mi 
        ({\hat{d}_{\vec{i}}-\hat{d}_{\vec{i}}^\dagger})/\sqrt{2}$. 
        In the bosonic case they are indeed
position and momentum operators fulfilling the 
canonical commutation
relations (CCR).
In turn, for fermionic operators, they are so called
{\it Majorana operators} fulfilling the canonical anti-commutation relations (CAR).
We will assume isotropic couplings for fermions, $C=0$, and
coupling only in position for bosons, $A-C=\id$. In order not
to obscure our main point, we will not consider the 
straightforward but cumbersome generalization to 
anisotropic or momentum couplings. The Hamiltonian now reads
($V_B:=A+C$, $V_F:=A$)
\begin{equation}
\label{Ham}
        \hat{H}_B =\frac{1}{2}
        \hat{\vec{r}}^t\left[\begin{array}{cc}
        V_B & 0\\
        0 & \id
        \end{array}\right]\hat{\vec{r}},\,
        \hat{H}_F=
        \frac{\mi}{2}\hat{\vec{r}}^t\left[\begin{array}{cc}
        0 & V_F\\
        -V_F^t & 0
        \end{array}\right]\hat{\vec{r}}
\end{equation}
for bosons and 
fermions, respectively.
Whenever we may treat both species equally, we 
denote by $V$ the coupling in position $V=V_B$ for 
bosons and $V=V_F$ for fermions.

Starting from the Hamiltonians above, their spectrum and respective ground 
states are found in the usual way by diagonalizing $\hat{H}_B$ through 
symplectic transformations, 
transformations respecting the CCR, and $\hat{H}_F$
by orthogonal transformations, 
transformations respecting the CAR.
As matrices $V$ are cyclic, the bosonic spectrum is given by 
$\lambda_{\vec{k}}$ and the fermionic spectrum by 
$|\lambda_{\vec{k}}|$, where 
$\lambda_{\vec{k}}=\sum_{\vec{l}\in\mathcal{L}}
V_{\vec{l}}\cos(2\pi\vec{k}\vec{l}/n)$, $\vec{k}\in \mathcal{L}$.

The half-space geometry allows for a 
transformation of both Hamiltonians to a system
of mutually uncoupled one-dimensional chains 
while respecting the CCR and CAR, but notably, 
changing the local
properties of the systems forming the 
individual chains. To this end consider the transformation 
$\hat{\vec{r}}=(\mathcal{O}\oplus\mathcal{O})\hat{\vec{q}}$ to a new set of operators $\hat{\vec{q}}$, 
where the $|\mathcal{L}|\times |\mathcal{L}|$ matrix $\mathcal{O}$ is given by 
$\mathcal{O}_{\vec{i},\vec{j}}=\delta_{i_1,j_1}O_{\vec{i}',\vec{j}'}$.
Here and in the following we write vectors $\vec{i}=(i_1\cdots i_D)\in\mathcal{L}$ as
$\vec{i}=(i_1\vec{i}')$, $\vec{i}'=(i_2\cdots i_D)\in\mathcal{L}'$.
Now, the transformation $\mathcal{O}$ acts on the first coordinate as the identity and is thus local with respect to the bipartition $\mathcal{A}|\mathcal{B}$, i.e., it does not change entanglement properties. In order to respect the CAR and CCR  
the matrix $O$ needs to be 
orthogonal. Then, the Hamiltonians $\hat{H}_{B/F}$ read in coordinates $\hat{\vec{q}}$ just as 
in Eq.~(\ref{Ham}) with modified coupling matrices 
$V\mapsto\mathcal{O}^tV\mathcal{O}$. As 
cyclic matrices commute and may be diagonalized by the same Fourier transformation, we have the explicit form 
\begin{equation}
        \nonumber        
        V_{\vec{l}}\!=\!\!\!
        \sum_{\vec{k}'\in \mathcal{L}'}\!\!\!
        \frac{\lambda_{\vec{k}'}\!(l_1)\me^{2\pi \mi \vec{k}'\!\cdot\vec{l}'/|\mathcal{L}'|}}{|\mathcal{L}'|},\,
        \lambda_{\vec{k}'}\!(l_1)\!=\!\!
        \sum_{k_1=1}^N\!\!
        \frac{\lambda_{(k_1\vec{k}')}\me^{2\pi\mi k_1l_1/N}}{N},
\end{equation}
where the $\lambda_{(k_1\vec{k}')}=\lambda_{\vec{k}}$, $\vec{k}\in\mathcal{L}$, 
are the eigenvalues of $V$. 
Now, define the $|\mathcal{L}'|\times |\mathcal{L}'|$ matrices $V\!(l_1)$
as those matrices obtained 
from $V$ by keeping the first coordinate fixed:  $(V\!(l_1))_{\vec{l}'}=V_{(l_1\vec{l}')}$. Then the $\lambda_{\vec{k}'}\!(l_1)$ are the eigenvalues of the $V\!(l_1)$, which are all cyclic and can thus all be
diagonalized by the same orthogonal matrix. Choosing $O$ to be this matrix yields
$(\mathcal{O}^tV\mathcal{O})_{(i_1\vec{k}'),(j_1\vec{p}')}=\delta_{\vec{k}',\vec{p}'}\lambda_{\vec{k}'}\!(i_1-j_1)$, a {\it momentum space} representation of the coupling in all
but the first coordinate. In this representation the Hamiltonian
is a sum of $|\mathcal{L}'|$ mutually uncoupled one-dimensional chains labeled by $\vec{k}'$. Each chain is described by a Hamiltonian of the form as in Eq.~(\ref{Ham}) with $N\times N$ cyclic coupling matrices $V\!(\vec{k}')$, $(V\!(\vec{k}'))_{l_1}=\lambda_{\vec{k}'}\!(l_1)$. We will write $(V\!(\vec{\varphi}'))_{\phi_1}=\lambda_{\vec{\varphi}'}\!(\varphi_1)$, $\phi_1=2\pi l_1/N$, $\varphi_d'=2\pi {k_d'}/{N}$, for the infinite system.
After this decoupling procedure the entanglement between $\mathcal{A}$ and $\mathcal{B}$ is now given by a sum of the entanglement
between the sites $[1,\dots,M]$ and $[M+1,\dots,N]$ of the individual chains:
\begin{equation}
\nonumber
\mathcal{E}=
        \lim_{N\rightarrow\infty}\sum_{\vec{k}'}
        \frac{E_S(\vec{k}')}{N^{D-1}}=\int_{[0,2\pi]^{\times D-1}}\!\!\!\!\!\!\!\!\!\!\!\!\!\!\!\!\!\! E_S(\vec{\varphi}')\frac{\md\vec{\varphi}'}{(2\pi)^{D-1}}.
\end{equation} 
This will be the 
starting point for the following discussion.

{\em Fermions. --} We start with investigating case (ii) above: The asymptotic 
behavior in $M$ after taking the limit 
$N\rightarrow\infty$ \cite{FermiDiscussion}, frequently
referred to as the {\it double scaling limit}  \cite{Keating}.
For each chain $\vec{\varphi}'$ we now need to compute the
entanglement between the first $M$ sites and the rest of the 
chain  (in real space, not momentum space). 
The asymptotic behavior in $M$ of this entanglement
can be obtained from the so-called {\it symbol} $g_{\vec{\varphi}'}\!(\varphi_1)=
        \sg(\lambda_{\vec{\varphi}'}\!(\varphi_1) )$ of the chain \cite{Keating,Singleshot}: Each $E_S({\vec{\varphi}'})$ is determined from the continuity properties of $g_{\vec{\varphi}'}$ as  a function of $\varphi_1$. For fixed $\vec{\varphi}'$ it corresponds to a 
one-dimensional isotropic fermionic model, for which 
the asymptotic form of the entanglement has been 
obtained in Refs.\ \cite{Keating},
\begin{equation}
\label{KOREPIN}
        E_S(\vec{\varphi}')=\frac{s(\vec{\varphi}')}{6}\log_2 M
        +c(\vec{\varphi}') +
        o(\log M ),
\end{equation}
where the function $c(\vec{\varphi}')$ does not depend on $M$ and 
the integer $s(\vec{\varphi}')$ is the number of discontinuities of 
$g_{\vec{\varphi}'}$ as function of $\varphi_1$
in the interval $[0,2\pi)$. All chains with $s(\vec{\varphi}')>0$ are critical as finding discontinuities in the symbol is equivalent to
having a vanishing energy gap above the ground state. This means that, 
depending on the Fermi-surface (the set of solutions to $\lambda_{\vec{\varphi}}=0$, see Fig.~\ref{fig}), one finds a continuum of chains $\vec{\varphi}'$ that are critical. This situation is in contrast to the situation encountered when considering bosonic systems as we will see below.
From (\ref{KOREPIN}), we find the asymptotic behavior in $M$ as
\begin{equation}
\label{ResultFermionic}
\mathcal{E}
        =\frac{\log_2M}{6}\sum_{\sigma=1}^\infty \frac{\sigma v(\Phi_\sigma)}{(2\pi)^{D-1}}
        +
        \int\frac{\md\vec{\varphi}'\,c(\vec{\varphi}')}{(2\pi)^{D-1}}+ o(\log M),
\end{equation}
where we defined $v(\Phi_\sigma)=\int_{\Phi_\sigma}\!\!\!\!\md\vec{\varphi}'$ as the volume of the 
set $\Phi_\sigma=\{\vec{\varphi}':
s(\vec{\varphi}')=\sigma\}$, so the set associated with exactly
$\sigma$ discontinuities.
Hence, we do encounter a {\it logarithmic divergence} in $M$ of 
the entanglement entropy and the pre-factor
depends on the topology of the Fermi-surface: $g_{\vec{\varphi}'}$
exhibits discontinuities at points where $\lambda_{\vec{\varphi}}=0$, i.e., on the Fermi-surface. If the Fermi surface is of measure zero (i.e., the
set of solutions to $\lambda_{\vec{\varphi}}=0$ is countable,
as for example in the critical bosonic case, see Fig.~\ref{fig}), we have
$v(\Phi_\sigma)=0$ and the system obeys the area law, $\mathcal{E}=\text{const}$. Consider as an example the case of
a nearest neighbor Hamiltonian with coupling 
$V_{\vec{i},\vec{j}}=\delta_{\vec{i},\vec{j}}+a\delta_{\text{dist}(\vec{i},\vec{j}),1}$, 
in which case the symbol corresponds for fixed $\vec{\varphi}'$ 
to that of the isotropic XY model with 
transverse magnetic field 
$h(\vec{\varphi}')\!\!=\!\!1+2a\sum_d\cos(\varphi_d')$. For
this model, the non-leading order term was obtained employing 
Fisher-Hartwig type methods \cite{Keating}. It reads
$c(\vec{\varphi}')=\log_2 (1- h^2(\vec{\varphi}') /4 )/6 +c_0$ 
and vanishes if the chain $\vec{\varphi}'$ is non-critical. 
The constant $c_0$ is independent of the system parameters.
The number of discontinuities is $s(\vec{\varphi}')=2$  for
\begin{equation}\nonumber
\vec{\varphi}'\in\Phi_2=\big\{\vec{\varphi}'\in [0,2\pi)^{D-1}:\big|\frac{1}{2a}+\sum_d\cos(\varphi_d')\big|< 1\big\}
\end{equation} 
and zero otherwise, i.e., the sum over $\sigma$ in (\ref{ResultFermionic}) consists only of the $\sigma=2$ term as all others are zero.
Thus
\begin{equation}\nonumber
        \mathcal{E}=
        \frac{v(\Phi_2)}{3(2\pi)^{D-1}}\log M 
        +\int_{\Phi_2}\frac{c(\vec{\varphi})\,\md\vec{\varphi}'}{(2\pi)^{D-1}}+o(\log M),
\end{equation}
where for $D=2$ and the critical case $|a|>1/4$,
we find $v(\Phi_2)=2\arccos\left(1/({2|a|})-1\right)$,
i.e., the prefactor depends on the coupling parameter $a$.
For non-critical models in the isotropic setting at hand 
the set $\Phi_2$ is empty and there is no entanglement. 
There is no universal non-leading order term as proposed 
in Ref.\ \cite{Preskill} related to the conformal charge, 
due to the specific geometric setting of a half-space.

At this point, it is interesting to discuss the behavior of the 
entanglement entropy under {\it Lifshitz phase transitions}. 
They are topological quantum phase transitions of fermionic 
systems due to a change of the topology of the Fermi surface, 
occurring for example in $d$-wave superconductors \cite{Vojta},
see also Ref.\ \cite{Roscilde}.
The previous considerations immediately allow us to argue that 
a Lifshitz transition accompanied by a change of topology 
of the Fermi surface is reflected by a non-analyticity in the 
prefactor $v(\Phi_\sigma)$ of the entanglement scaling law: Any 
change of the topology will lead to a non-differentiable
alteration of the prefactor of the leading order term.

The second setting is the one of $M=N/2$ (case (i) above), 
where---in contrast to the double scaling limit---$M$ depends on $N$, 
leaving one limit to consider. From the discussion above, we would expect $\mathcal{E}=\infty$. Indeed, making use of a quadratic lower bound to the 
entanglement entropy \cite{Fannes,Wolf}, it can be shown that
fermions violate the area law in this setting: 
For fermionic models with nearest-neighbor interactions with 
half-filling in $D=2$, so 
$V_{\vec{i},\vec{j}}=a\delta_{\text{dist}(\vec{i},\vec{j}),1}$, we find after some straightforward 
algebra \cite{Alg} 
$ \lim_{N\rightarrow\infty}\sum_{k=1}^NE_S(k)/N \geq 
\lim_{N\rightarrow\infty} \sum_{l=1}^{N/2-1}\!\!4/(\pi^2l)=\infty.$

{\em Bosons. --} We concentrate on the geometrical setting $M=N/2$ (case (i) above) and the most
significant model: The $D$-dimensional free 
{\it Klein Gordon field}, 
\begin{equation}
\label{KG}
\hat{H}=\frac{1}{2}\int_{[0,L]^{\times D}}\!\!\!\!\!\!\!\!\!\md\vec{r}\left[\pi(\vec{r})^2+\mu^2\phi(\vec{r})^2
+v^2\left(\vec{\nabla}\phi(\vec{r})\right)^2\right],
\end{equation}
which may be obtained from a Hamiltonian as in Eq.~(\ref{Ham})
with a nearest  neighbor coupling 
$V_{\vec{i},\vec{j}}=(\mu^2+2D\Omega^2)\delta_{\vec{i},\vec{j}}-
\Omega^2\delta_{\text{dist}(\vec{i},\vec{j}),1}$:
Denoting the lattice spacing by $\alpha=L/N$ and taking the limit $N\rightarrow\infty$ while keeping $v^2=\Omega^2\alpha^2$ constant, one obtains (\ref{KG}).
Rescaling w.l.o.g. $V\mapsto V/(\mu^2+2D\Omega^2)$, we find
$V_{\vec{i},\vec{j}}=\delta_{\vec{i},\vec{j}}-
c_N\delta_{\text{dist}(\vec{i},\vec{j}),1}$ and 
$ c_N=(\mu^2L^2/(v^2N^2)+2D)^{-1}\;\;\substack{\longrightarrow \\ N\rightarrow\infty}\;\;1/2D
$ .
Now, demanding the system to be critical uniquely determines 
$c_N\rightarrow 1/(2D)$ as the energy gap between the 
ground and first excited state is given by the square root of the smallest eigenvalue of the coupling matrix $V$.

For each individual chain $\vec{k}'$, the $V\!(\vec{k}')$ are 
transformed nearest-neighbor coupling matrices:
\begin{equation}\nonumber
        \left(V\!(\vec{k}')\right)_{i,j}\!\!=
        \big[1-2c_N
        \sum_d\cos(2\pi k'_d/N)\big]\delta_{i,j}
        -c_N\delta_{\text{dist}(i,j),1}.
\end{equation}
In the infinite system limit, the energy gap $\Delta E(\vec{\varphi}')$ between the ground and first excited state
of each chain is given by $\Delta E(\vec{\varphi}')^2D=D-
\sum_d\cos(\varphi_d')-1$. Hence, in contrast to the critical fermionic case, the set of solutions to $\lambda_{\vec{\varphi}}=0$ is of measure zero as only a single chain becomes critical, when we 
identify $0$ with $2\pi$ in the spectrum, 
see Fig.~\ref{fig}. We now make use of a powerful result of 
Ref.\ \cite{Harmonic}: The exact form of the logarithmic 
negativity $E_n$ (an upper bound to the entropy of entanglement 
and an entanglement monotone \cite{Vidal,Eisert})
with respect to the split $[1,\dots,N/2]|[N/2+1,\dots,N]$
for a harmonic chain with nearest neighbor coupling. 
It is not only an asymptotic statement in $N$, but a
closed-form expression and we find that $\lim_{N\rightarrow\infty}\sum_{\vec{k}'}E_n(\vec{k}')/N^{D-1}$ converges to
\begin{equation}\nonumber
\frac{1}{2(2\pi)^{D-1}}\int_{[0,2\pi]^{D-1}}\!\!\!\!\!\!\!\!\!\!\!\!\!\!\!\!\md\vec{\varphi}'\,\log_2
        \Bigl(\frac{D-\sum_{d=1}^{D-1}
        \cos(\varphi_d')+1}{D-\sum_{d=1}^{D-1}
        \cos(\varphi_d')-1}
        \Bigr)\ge\mathcal{E},
\end{equation}
independent of the mass $\mu$.
For $D=2$ it evaluates to $\log_2(3+2\sqrt{2})/2$
and similarly for $D>2$. Hence, the entanglement entropy for this critical model is bounded by an 
expression linear in the {\it boundary area}, we
do not encounter an infrared divergence here, and 
the prefactor can be exactly determined
 for the logarithmic negativity. 
 
{\em Summary. --} In this work, we have clarified the issue of scaling of the 
entanglement entropy
in bosonic and fermionic lattice systems. Our analytical argument
indeed confirms and resolves 
previous numerical findings and conjectures on the scaling of 
entanglement in ground states of many-body systems. 
The difference between the behavior of bosons and 
fermions may be taken as unexpected. 
A new length scale emerges
that can be referred to as ``entanglement
thickness'': This length scale is associated with the possibility of approximately
disentangling the two regions with local operations. As a pure Gaussian state
is equivalent to a product of two-mode squeezed states up to local unitary rotations, one finds
that this scale is in fact associated with the lattice spacing, and not the
correlation length. The 
violation of the area-law for fermions is in fact 
intertwined with the specific role of the Fermi surface. 
We found that quantum phase transitions involving an alteration 
of the topology of the Fermi surface result in a non-analytical 
behavior of the prefactor.  

 For typical critical models we found that 
those individual boundary crossing chains 
obtained from our decoupling procedure that are 
critical form a continuum for fermions and 
are finite in number for bosons. Integrating over all chains then 
"lifts" singularities in $D>1$ for bosons. 
For critical bosonic models exhibiting only a finite 
number of ground states, one might expect that 
an area law holds in dimensions $D>1$ 
even for models that go beyond quadratic 
Hamiltonians considered here and are thus truly 
interacting. Confirming or refuting this conjecture
is an interesting challenge.

{\it Acknowledgements. --}
This work has benefited from discussions with
 M.M.\ Wolf, T.J.\ Osborne, M.\ Ericsson, and M.B.\ Hastings.
It has been supported by the DFG
(SPP 1116), the EU (QAP), the Royal Society, 
the QIP-IRC, 
Microsoft Research, 
and the EURYI Award Scheme.

\end{document}